\documentstyle[11pt,paspconf,epsf]{article}

\begin{document}

\rightline{To be published in the {\it Proceedings of IAU Symposium 191: AGB Stars}}

\title{  Link between Mass-loss and Variability Type for AGB Stars?   } 

\author{        \v{Z}eljko Ivezi\'{c},  Gillian R. Knapp      }

\affil{                 Princeton University,
                  Department of Astrophysical Sciences,
                     Princeton, NJ 08544-1001, USA                    
}

\begin{abstract}

We find that AGB stars separate in the 25-12 vs. 12-K color-color
diagram according to their chemistry (O, S vs. C) and variability type
(Miras vs. SRb/Lb).  While discrimination according to the chemical
composition is not surprising, the separation of Miras from SRb/Lb
variables is unexpected.

We show that ``standard'' steady-state radiatively driven models
provide excellent fits to the color distribution of Miras of all chemical
types. However, these models are incapable of explaining the dust
emission from O-rich SRb/Lb stars. The models can be altered to
fit the data by postulating different optical properties for silicate
grains, or by assuming that the dust temperature at the inner envelope
radius is significantly lower (300-400 K) than typical condensation
temperatures (800-1000 K), a possibility which is also supported by
the detailed characteristics of LRS data.  While such lower temperatures 
are required only for O- and S-rich SRb/Lb stars, they are also consistent 
with the colors of C-rich SRb/Lb stars.

The absence of hot dust for SRb/Lb stars can be interpreted as a recent
(order of 100 yr) decrease in the mass-loss rate. The distribution of
O-rich SRb/Lb stars in the 25-12 vs. K-12 color-color diagram shows
that the mass-loss rate probably resumes again, on similar time scales.
It cannot be ruled out that the mass-loss rate is changing periodically
on such time scales, implying that the stars might oscillate between
the Mira and SRb/Lb phases during their AGB evolution as proposed by
Kerschbaum et al. (1996). Such a possibility appears to be supported by
recent HST images of the Egg Nebula obtained by Sahai et al. (1997),
the discovery of multiple CO winds reported by Knapp et al. (1998), and
long-term visual light-curve changes detected for some stars by Mattei
(1998).

\end{abstract}


\section{Introduction}

AGB stars are long-period variables (LPV) and show a variety of light
curves. Based on visual light curves, the General Catalog of Variable
Stars (GCVS) defines regular variables, or Miras, semiregular (SR)
variables, and irregular variables (L). The distinctive features are
the regularity of light curves, their amplitude and their period. Some 
types are further subdivided based on similar criteria (e.g. SRa, SRb,
SRc,...).

AGB stars of different variability types cannot be distinguished by
considering IRAS PSC data alone.  Kerschbaum, Hron and collaborators
(Kerschbaum \& Hron 1996, and references therein; hereafter KH) have
studied IR emission of various types of LPVs by combining IRAS
broad-band fluxes with near-IR observations and spectra from the IRAS
LRS database. They find that SRa variables appear to be a mixture of 
two distinct types:
Miras and SRb variables. SRb variables have somewhat higher stellar
temperatures and smaller optical depths than Miras, and can be further
divided into ``blue" and ``red" subtypes, the former showing much less
evidence for dust emission than the latter.

Particularly intriguing are differences between the sLRS spectra of SRb
variables and Miras with ``10" $\mu{\rm m}$ silicate emission feature
(LRS class 2n). The peak position of the ``10" $\mu{\rm m}$ feature for
SRb sources is shifted longwards, relative to the peak position for
Miras, by 0.2-0.3 $\mu{\rm m}$. In addition, the ratio of the
strengths of silicate emission features at 18 $\mu{\rm m}$ and 10
$\mu{\rm m}$, $F_{18}/F_{10}$, is larger for SRb variables than for
Miras.  These results are confirmed at high statistical significance
by Marengo (1998), who determines the ``10" $\mu{\rm m}$ feature
peak position by fitting a polynomial to the 9-11 $\mu{\rm m}$ spectral
region.  He finds a difference in median peak positions of 0.2 $\mu{\rm
m}$, with probability larger than 0.9999 that the two distributions are
different.  

Motivated by these results, we further analyze correlations between IR
emission and variability type for AGB stars, and their implications for
the models of steady-state radiatively driven outflows. We base our
analysis on IRAS data and a set of JHKLM photometric observations for
about 600 LPVs, kindly made available to us by Franz Kerschbaum.

\section{Distribution of AGB stars in K-12-25 Color-color Diagram}

Following the results of KH, we consider only SRb, Lb, and Mira
variables, and further divide them according to grain chemistry.  Fig.
1 displays the distribution of sources in the K-12-25 diagram \footnote{We
use colors defined as $\lambda_2 - \lambda_1 = \log(F_\nu(\lambda_2) /
F_\nu(\lambda_1))$, where $F_\nu(\lambda_1)$ and $F_\nu(\lambda_2)$ are
fluxes at wavelengths $\lambda_1$ and $\lambda_2$.}.

\begin{figure}
\plotone{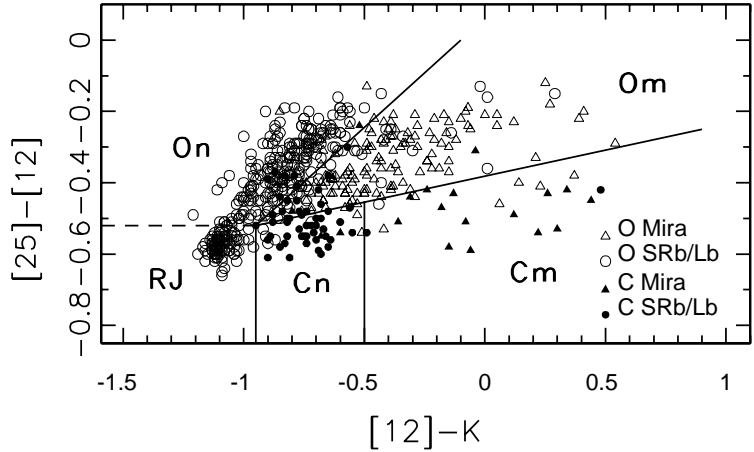}
\caption{K-12-25 color-color diagram for mass losing AGB stars.
Various symbols correspond to source variability type and grain 
chemistry, as marked. Solid lines display classification 
into 4 regions according to grain chemistry (O and S stars with silicate 
grains vs. C stars with carbonaceous grains), and variability type 
(Miras, m, vs. non-Miras, n). Dashed line separates On sources near
the Rayleigh-Jeans point (RJ) which have no, or very little dust.}
\end{figure}

Stars in our sample clearly separate according to grain chemistry (O
vs. C stars, S stars cannot be distinguished from O stars). Such a
clear separation is not very surprising since the optical properties of
silicate and carbon grains, which by and large determine the spectral
shape of dust emission, are significantly different at these 
wavelengths.  Stars also separate according to their GCVS variability
type into two well defined regions, Miras and non-Miras (i.e. SRb and Lb,
which exhibit similar behavior). This is in sharp contrast with the 
distribution of the same sample in the IRAS 12-25-60 diagram, which shows 
no such division.

A scheme which classifies stars
into 4 groups according to their variability type (Miras, m, vs. 
non-Miras, n) and grain chemistry (O stars with silicate grains
and C stars with carbonaceous grains) is shown in Fig. 1 by solid lines. 
The boundaries, required to be straight lines, are determined by maximizing 
the efficiency and reliability of the resulting classification. 
The dashed line separates On sources around the Rayleigh-Jeans point
and is determined by assuming a comparable scatter in both
colors. 


\section{Interpretations of the Source Distribution in
		K-12-25 Color-color Diagram}

\subsection { Predictions of the ``Standard" Models}

We first attempt to model the source distribution by employing spectra
expected for a steady-state radiatively driven outflow with either
silicate or amorphous carbon grains (Ivezi\'c \& Elitzur 1995, and
references therein). We define ``standard" models by 2500 K black body
stellar spectral shape, and dust temperature at the inner 
envelope edge of 1200 K for carbon grains and 700 K for silicate grains.
These two temperatures mark the ``standard" model tracks in Fig. 2. For 
each type of grain, the model track starts at the Reyleigh-Jeans point, 
with the positions along the track parametrized by dust optical depth. 

\begin{figure}
\plotone{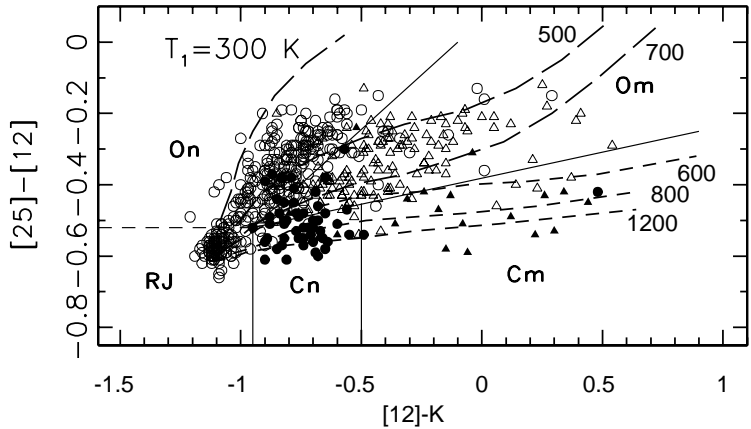}
\caption{K-12-25 color-color diagram for mass losing AGB stars.
Symbols correspond to source variability type and grain chemistry, as
marked in Fig. 1. Long-dashed lines are model tracks for silicate grains with
various dust temperatures at the inner envelope radius, T$_1$, as
marked in the figure. Models for amorphous carbon grains
are displayed by short-dashed lines. ``Standard" models have T$_1$=700 K
(silicate grains) and T$_1$=1200 K (carbon grains).}
\end{figure} 

The track for carbon grains passes through the distribution
of stars with carbon dust, i.e. through the regions Cn and Cm. The
scatter of points is marginaly consistent with the photometric errors,
and non-Mira stars appear to be smaller optical depth counterparts of
Miras.  The model track for silicate dust passes through the
distribution of Miras in Om region, although not through the region
with the highest density of sources. The offset of about 0.5 mag (0.2
on plotted log scale) is somewhat large, but still not sufficient to
clearly rule out the model. However, the model prediction is in dire
disagreement with the distribution of non-Miras with silicate dust
(region On). For example, for 25-12 color of -0.3, the discrepancy in
12-K color is about 0.75 on log scale, or almost 2 magnitudes.

\subsection{ Variations of the ``Standard" Models }

There are several ways to augment the ``standard" models without 
significantly changing the paradigm of steady-state radiatively
driven outflow.

Stellar spectral shape can effect the model spectrum when dust optical
depth is not large. For example, increasing the stellar temperature 
moves a model track to the left in the K-12-25 color-color diagram, as required
by the data. However, we are unable to produce satisfactory model tracks 
by varying the stellar temperature within a plausible range ($<$ 4000 K).
Another possibility is that a black-body spectral shape is not
an adequate description of the true stellar spectrum. This can be
ruled out by considering the K-L-M color-color diagram: differences in median 
near-IR colors between Miras and non-Miras are about 3-4 times smaller 
than required to account for the observed differences in 12-K color.
Such small differences in near-IR data also rule out an absorption feature 
in the K band, e.g. H$_2$O feature as seen in the data by Matsuura et al. 
(1998), as an explanation for redder 12-K colors observed for Miras.
Furthermore, similar separation of stars into 4 regions can also be 
seen in the L-12-25 color-color diagram.

Models in agreement with data for non-Miras can be produced by altering
the spectral shape of absorption efficiency for silicate grains. There
are two required changes: the ratio of the strengths of silicate
features at 18 $\mu{\rm m}$ and 10 $\mu{\rm m}$ has to be increased by a
factor of about 2-3, and the peak position of the ``10" micron feature
has to be shifted longwards by about 0.2-0.3 $\mu{\rm m}$.  Such 
absorption efficiency results in a model track which leaves the 
Rayleigh-Jeans point at a larger angle than before, and passes exactly 
through the observed source distribution.

Another way to make ``standard" models agree with the data is to change
T$_1$, the assumed dust temperature at the inner envelope edge.  The
model tracks for (unaltered) silicate grains with T$_1$ = 300, 500 and
700 K are shown in Fig. 2 by solid lines. They demonstrate that the
distribution of stars with silicate dust in the K-12-25 color-color diagram
can be reproduced by varying T$_1$ in the range 300--700 K, without any
change in the adopted absorption efficiency for silicate grains.  It is
remarkable that these models are also capable of explaining the
differences between Miras and non-Miras found in LRS data.  Fig. 3
displays two model spectra obtained with visual optical depth of 0.6,
and T$_1$ = 300 K (dashed line) and 700 K (solid line). The dotted line
shows the stellar spectrum.  Because the hot dust has been removed in
the model with T$_1$ = 300 K, both the blue and red edges of the silicate 10
$\mu{\rm m}$ emission feature are shifted longwards by about 0.2-0.3
$\mu{\rm m}$, causing the whole feature to appear shifted although the peak 
position is the same in both models. In addition, this model has a larger 
$F_{18}/F_{10}$ flux ratio, as observed. This success 
and the application of Occam's razor hints that the absence of hot
dust is the most likely explanation for differences in IR spectra
between Mira and non-Mira stars with silicate dust.

\begin{figure}
\plotone{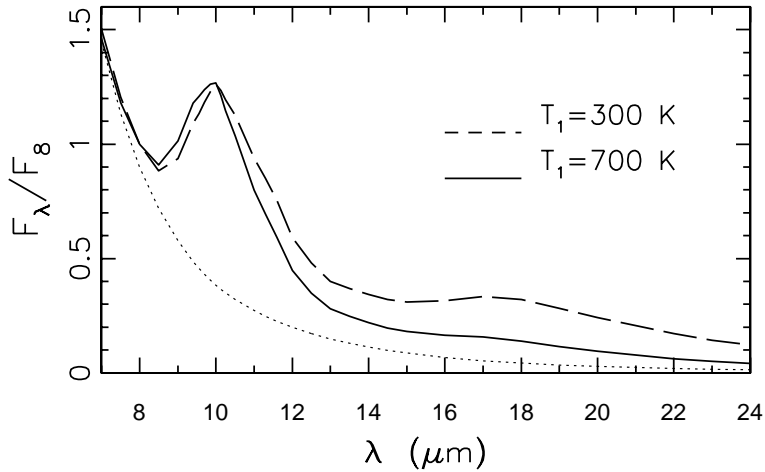}
\caption{Model spectra in the LRS region obtained for silicate grains
and 2 values of the dust temperature at the inner envelope radius,
T$_1$ (solid and dashed lines). The dotted line shows the stellar 
contribution to the spectra. Note the horizontal shift between the
features, although both are calculated with the same grain optical 
properties.}
\end{figure} 


The distribution of Miras and non-Miras with carbon dust is fairly
well described by the ``standard" model with T$_1$ = 1200 K, and
thus the assumption of significantly lower T$_1$ is not required by
the data. However, such models are not ruled out by the data as
shown by model tracks for T$_1$ = 600, 800 and 1200 K displayed
by dashed lines in Fig. 2.

\section{Disussion}

Absence of hot dust, as implied by low T$_1$, might be a natural 
consequence of a significant recent decrease in mass-loss rate.
From the distribution of sources in the On region, we infer a time 
scale of about 100 years. What happens after that?
The distribution of On sources ends quite sharply at its blue edge. 
If envelopes continued to freely expand, distribution would
extend smoothly to bluer 12-K, and redder 25-12 colors. The absence
of these sources argues that mass-loss resumes after about 100 years.
This idea, that stars oscillate between the SRb and Mira phases,
was proposed by Kerschbaum et al. (1996) based on similar number
densities and scale heights.

This possibility seems to be supported by a number of new observations.
Recent HST images of the Egg Nebula obtained by Sahai et al. (1997)
show concentric shells whose spacing corresponds to a dynamical time
scale of about 100 years. The discovery of multiple CO winds reported
by Knapp et al. (1998) is in agreement with the hypothesis of variable
mass-loss. Furthermore, they find that the velocity and mass-loss rate
for each individual component follows the relation $v \propto
\dot{M}^{1/3}$ found for stars with single CO winds (Young, 1955), and
theoretically expected for radiatively driven outflows (Ivezi\' c,
Knapp \& Elitzur, 1998). Finally, there is mounting direct evidence
that long-term visual light curve changes are consistent with the
transition between Mira and non-Mira variability types (Mattei, 1998).

If proven correct, the mass-loss variations on time scales of about 100
years would significantly change our understanding of the stellar
evolution on the AGB. What type of observations could rule out this
hypothesis? The most obvious and direct test is mid-IR imaging.  When
T$_1$ decreases by a factor of two, the inner envelope radius
increases by about a factor of four. Such a difference should be easily
discernible for a few dozen candidate stars with largest angular sizes,
which are already within the reach of the Keck telescopes (Monnier, 1998).

\acknowledgments

We are grateful to Franz Kerschbaum for his JHKLM observations
without which this work would not have been possible. Massimo Marengo
and Mikako Matsuura have endebted us by communicating their results 
prior to publication. This work was supported by NSF grant AST96-18503.

\end{document}